# Single-cell phase-contrast tomograms data encoded by 3D Zernike descriptors

Pasquale Memmolo,[1,*] Daniele Pirone,[1] Daniele G. Sirico,[1] Lisa Miccio,[1] Vittorio Bianco,[1] Ahmed B. Ayoub,[2] Demetri Psaltis,[2] Pietro Ferraro[1]

[1] CNR-ISASI, Institute of Applied Sciences and Intelligent Systems "E. Caianiello", Via Campi Flegrei 34, 80078 Pozzuoli, Napoli, Italy.
[2] EPFL, Ecole Polytechnique Fédérale de Lausanne, Optics Laboratory, CH-1015 Lausanne, Switzerland.

* pasquale.memmolo@isasi.cnr.it

## Abstract

Phase-contrast tomographic flow cytometry combines quantitative 3D analysis of unstained single cells and high-throughput. A crucial issue of this method is the storage and management of the huge amount of 3D tomographic data. Here we show an effective quasi lossless compression of tomograms data through 3D Zernike descriptors, unlocking data management tasks and computational pipelines that were unattainable until now.

## Introduction

Tomographic Phase Microscopy (TPM) is the most innovative 3D microscopy technique for label-free imaging of single cells, providing rich information content[1-3]. The tomogram of a cell reproduces the refractive index (RI) distribution that intrinsically encodes the whole 3D intracellular structure. This opened the route for a comprehensive biophysical characterization of cells through their 3D-RI distribution that can be considered as an endogenous biomarker. In the last decade, very promising results of TPM in biomedicine have demonstrated the effectiveness of this technology as a novel imaging modality for single-cell studies[4,5]. The recent demonstration of TPM in flow cytometry condition, namely Tomographic Flow Cytometry (TFC), added the missing milestone for employing this technology in high-throughput modality[6,7]. The TFC microscope commonly uses digital holography to image rotating cells flowing along microfluidic channels[6]. Depending on the number and the velocity of cells simultaneously imaged in the Field of View (FoV), the throughput can vary from tens to thousands of cells per minutes. While this is a unique opportunity for in-depth single-cell analysis with high statistical relevance, it also poses a non-negligible problem in terms of data management. The huge amount of data has pushed the scientific community to improve computational processing in terms of costs and speed through machine learning[8]. The latest result reports the use of deep learning to make the holographic reconstruction process up to 45 times faster[9]. Learning approaches have been successfully employed to improve the quality of tomographic reconstructions[10-12] even in the case of very limited angle views[13,14]. Recently, it has been demonstrated the possibility to achieve potentially more than 10000 tomograms per second, using only 4 views acquired simultaneously with an angle-multiplexing illumination strategy and exploiting the deep learning to recover the missing information[13]. It is evident that, storing and managing 3D data in a fast and accessible way will be necessary to make TFC technology truly exploitable in clinical applications. Here we introduce, for the first time in the field of TPM, the use of a Zernike polynomials basis in 3D for reconstructing single-cell tomograms. We demonstrate the possibility to encode 3D data into a sequence (i.e. a 1D numerical string) of 3D Zernike Descriptors (3DZD) coefficients, without significant loss of information. The 3DZD encoding was demonstrated for 3D shape retrieval[15] and, very recently, it has been employed to identify the binding site for a specific ligand of G protein-coupled receptors[16]. To the best of our knowledge, the 3DZD representation of single cells tomograms has not been investigated before, despite the strong relation between Zernike polynomials and quantitative phase imaging of cells. In fact, few years ago, it was shown that cells can be modeled as opto-fluidic microlenses and 2D Zernike polynomials were used to characterize the optical aberrations of cells from their phase-contrast images[17]. Intuitively, the extension of this concept to 3D is immediate. Since the cell's phase-contrast image is the 2D projection of the 3D real object (i.e. the tomogram), the 3D extension of Zernike polynomials can be considered as the generalization of the microlens volumetric shaping. This concept opens to a new paradigm for managing tomograms due to the property of the 3D Zernike polynomials to be orthogonal basis functions. Hence the corresponding weights (i.e. 3DZD) encode tomographic data, thereby allowing quasi lossless compression.

## Results

Figure 1 reports the demonstration of the ability of 3DZD to reproduce tomographic data with very high fidelity. In particular, a tomographic cell phantom of sizes $50 \times 50 \times 50$ has been simulated (Fig. 1a) and fitted by using 3D Zernike polynomials up to order 30 (Fig. 1b), thus recovering the corresponding 3DZD (Fig. 1c). These have been used to reconstruct the tomogram (Fig. 1d). We used the Normalized Root Mean Square Error (NRMSE) to evaluate its similarity in comparison to the ground truth of Fig.1a, resulting in NRMSE = 0.81%. Finally, the Fig. 1e shows the trend of the tomogram recovery fidelity in terms of NRMSE using the 3DZD when the order of Zernike polynomials employed for the fitting grows. Notice that, tiny details (i.e. substructures with smaller sizes) become visible for high order Zernike polynomials (greater than 20) as highlighted by the inset isolevel images in Fig. 1e.

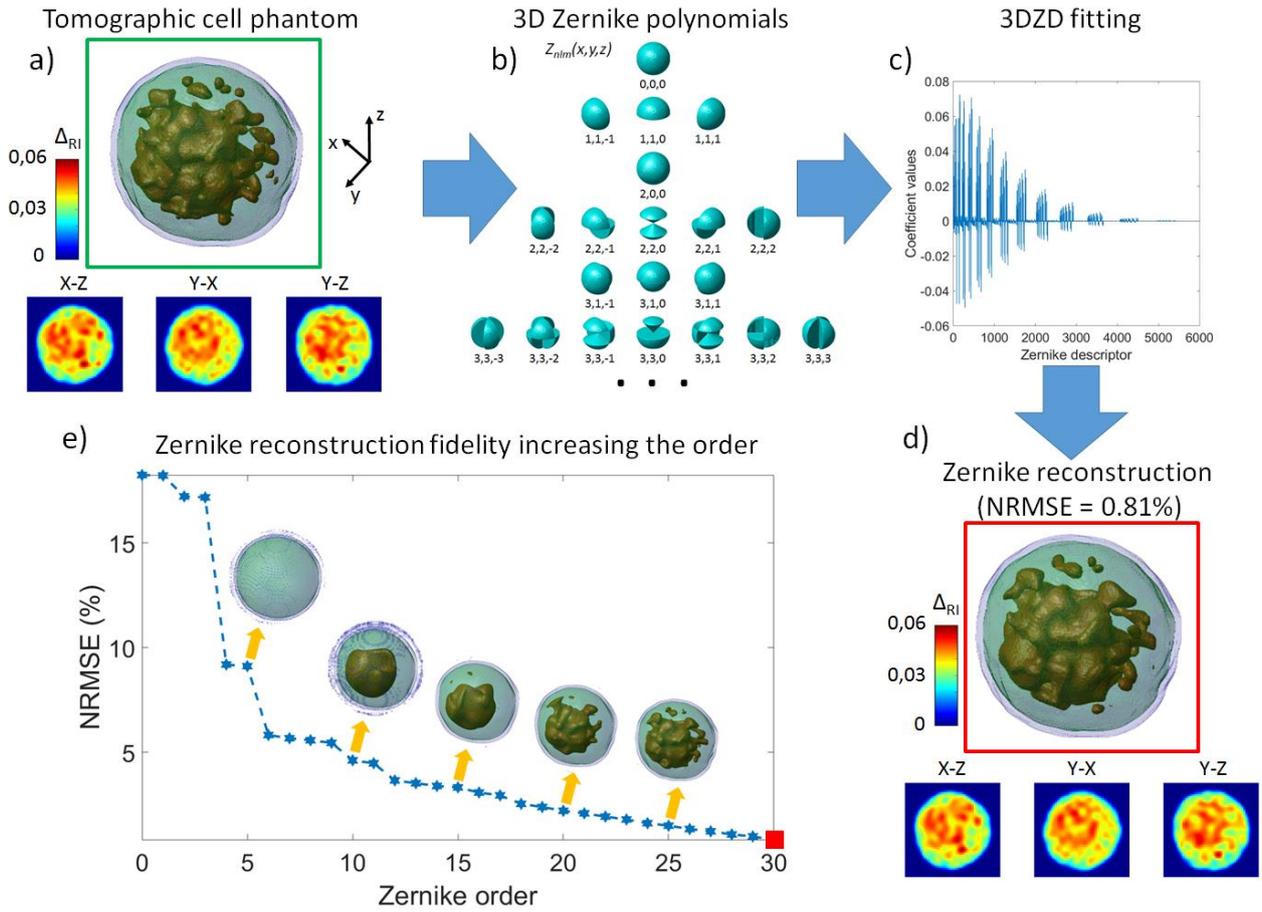

**Fig. 1. Proof of 3DZD encoding for high fidelity recovery of tomographic data.** (**a**) Isolevel image of Tomographic cell phantom with three inset images reporting the central slices along orthogonal directions. (**b**) Shapes visualization of the first 20 Zernike polynomials (up to order 3). (**c**) 3DZD obtained by fitting the tomographic cell phantom with Zernike polynomials up to order 30 (corresponding to 5456 descriptors). (**d**) Tomogram recovery by using the 3DZD, reporting a fidelity score NRMSE = 0.81%. The retrieved isolevel image and relative central slices show high visualization fidelity too if compared to (a). (**e**) NRMSE vs Zernike order. Inset figures report isolevel images reconstructed by fixing the Zernike polynomials order up to 5, 10, 15, 20, 25, showing the retrieval of tiny details when the order of basis functions employed for fitting grows.

The mathematical derivation of Zernike basis functions in 3D and the computation of the 3DZD are described in the Methods section. The validation of the proposed approach is achieved in case of experimental data on living cells, acquired using different tomographic imaging systems and processed by different tomogram reconstruction algorithms (see Fig. 2). Specifically, the red blood cell (Fig. 2a) was recorded by the imaging apparatus proposed in ref. [6], the white blood cell (Fig. 2b), the SKOV3 ovarian cancer cell (Fig. 2c), the CHP212 neuroblastoma cancer cell (Fig. 2d), and the NIH 3T3 mouse cell (Fig. 2e) were recorded by using the tomographic imaging system in ref. [9]. All these tomograms are obtained using the Filtered Back Projection (FBP) algorithm.

Finally the Yeast cell (Fig. 2f) was acquired in the illumination scanning based tomographic setup reported in ref. [18], while the corresponding tomogram reconstruction is performed by using the Learning Tomography (LT) method[19].

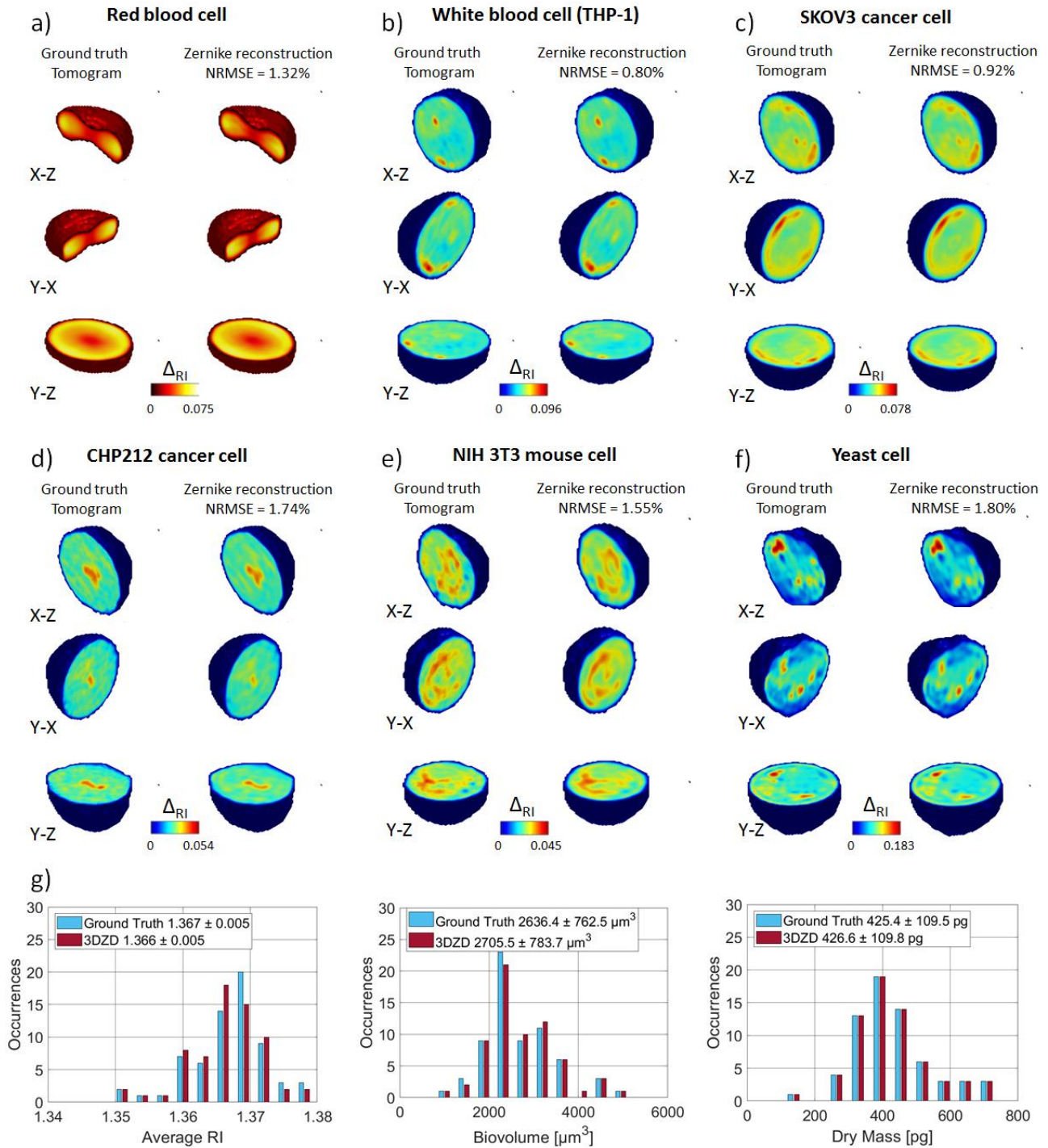

**Fig. 2. Comparison between experimental tomographic data and the corresponding 3DZD reconstructions.**
**(a-f)** Left side: central slices of tomographic reconstructions of cells from experimental data; right side: central slices of tomograms recovered via 3DZD. **(g)** Histograms of average RI, biovolume and dry-mass, calculated from both experimental tomograms and corresponding reconstructions via 3DZD, in the case of NIH 3T3 cell line.

More details about both tomographic imaging systems and reconstruction algorithms are reported in the Methods section. The corresponding tomograms, recovered from 3DZD, are shown on the right sides of Figs. 2 a-f, reporting a NRMSE < 2% in all cases. Notice that, if one would ensure the achievement of a desired fidelity for each cell in Fig.2, e.g. NRMSE < 1%, it is sufficient to further increase the order of the Zernike polynomials. A deeper performance analysis is included in the Methods section. The achieved high fidelity in recovering single-cell tomograms via 3DZD allows us to compute the most common biophysical features, such as biovolume, average RI and dry-mass, thus preserving the distinctive signature of each individual cell when compared to those obtained from experimental data. Fig.2g reports the comparison of histograms of the abovementioned features, calculated from both ground truth tomograms and 3DZD reconstructions, for the NIH-3T3 cell line.

## Discussion and conclusions

In summary, we demonstrated, for the first time in the field of tomographic phase microscopy, the possibility to encode single-cell tomograms by using the 3D version of Zernike polynomials. We proved a fidelity criterion based on the NRMSE in a way to reconstruct tomographic data of sizes $50 \times 50 \times 50$ from a sequence of 5456 Zernike descriptors achieving data compression ratio ~ 22.9 and a corresponding space saving > 95% in quasi lossless compression modality, i.e. NRMSE < 1%. These results open the route to a potentially breakthrough methodology to store, manipulate and process volumetric images. Moreover, the possibility to replace volumetric data with the corresponding 3DZD sequence allows to address clinical challenges related to label-free single-cell phenotyping, for which the current solution is based on deep learning classification[18,19]. In principle, by the proposed method, it would be possible to switch from deep 3D image classification to deep 1D sequence classification, preserving the same degree of data richness, i.e. with negligible information loss, and allowing a very remarkable memory and computational time reduction for the training of neural networks. Finally, the possibility to employ 3DZD as morphological biomarkers to identify specific subcellular structures may be achieved as a direct generalization from the 2D case, in which the Zernike fitting is used to characterize aberrations of cells as in ref. [17].

# Methods

## *Mathematical derivation of Zernike basis functions in 3D*

The 3D Zernike functions can be defined in spherical coordinates as[15]

$$Z_{nl}^m(r,\vartheta,\varphi) = R_{nl}(r) Y_l^m(\vartheta,\varphi), \qquad (1)$$

where $n, l, m$ are integer indexes such that $n \geq 0$, $l \leq n$ with $n - l$ even numbers and $m \in [-l, l]$. The radial part $R_{nl}(r)$ of Eq. (1) is defined as

$$R_{nl}(r) = r^l \sum_{v=0}^{k} q_{kl}^v \, r^{2v}, \qquad (2)$$

where $k = \frac{n-l}{2}$ and the coefficients $q_{kl}^v$ are determined so that the resulting basis functions are orthonormal in the unit ball

$$q_{kl}^v = \frac{(-1)^k}{2^{2k}} \sqrt{\frac{2l+4k+3}{3}} \binom{2k}{k} (-1)^v \frac{\binom{k}{v}\binom{2(k+l+v)+1}{2k}}{\binom{k+l+v}{k}}. \qquad (3)$$

The angular part $Y_l^m(\vartheta, \varphi)$ is set as the spherical harmonics, given by

$$Y_l^m(\vartheta, \varphi) = N_l^m P_l^m(\cos\vartheta) \, e^{im\varphi}, \qquad (4)$$

where $P_l^m$ denotes the associated Legendre functions and $N_l^m$ is a normalization factor defined as

$$N_l^m = \sqrt{\frac{(2l+1)}{4\pi} \frac{(l-m)!}{(l+m)!}}. \qquad (5)$$

In order to formulate the 3D Zernike polynomials as homogenous polynomials in the Cartesian coordinates $\mathbf{x} = (x, y, z)^T$, it is needed to rewrite spherical harmonics in harmonic polynomials. By using the conversion between Cartesian and spherical coordinates and the integral formula for associated Legendre functions, one can express the harmonic polynomials as follow

$$h_l^m(\mathbf{x}) = r^l Y_l^m(\vartheta, \varphi) = c_l^m \left(\frac{ix-y}{2}\right)^m z^{l-m} \sum_{\mu=0}^{\lfloor\frac{l-m}{2}\rfloor} \binom{l}{\mu}\binom{l-\mu}{m-\mu}\left(-\frac{x^2+y^2}{4z^2}\right)^\mu, \qquad (6)$$

where $c_l^m$ are normalization factors defined as

$$c_l^m = c_l^{-m} = \frac{\sqrt{(2l+1)(l-m)!(l+m)!}}{l!}. \qquad (7)$$

The Eq. (6) yields homogenous polynomials for $m > 0$. For $m < 0$ the following symmetry relation is used

$$h_l^{-m}(\mathbf{x}) = (-1)^m \overline{h_l^m(\mathbf{x})}. \qquad (8)$$

Therefore, the Eq. (1) can be rewritten in Cartesian coordinates

$$Z_{nl}^m(\mathbf{x}) = \sum_{v=0}^{k} q_{kl}^v \, |\mathbf{x}|^{2v} \, h_l^m(\mathbf{x}). \qquad (9)$$

*Computation of 3D Zernike descriptors*

To implement the Eq. (9), one can derive a more compact formulation by expanding $Z_{nl}^m(\mathbf{x})$ through the Eq. (6). After some mathematical manipulations (see details in ref. [15]), the Eq. (9) can be rewritten as

$$Z_{nl}^m(\mathbf{x}) = \sum_{w+s+t \leq n} \chi_{nlm}^{wst} x^w y^s z^t, \qquad (10)$$

where $\chi_{nlm}^{wst}$ is set as

$$\chi_{nlm}^{wst} = c_l^m 2^{-m} \sum_{v=0}^{k} q_{kl}^v \sum_{\alpha=0}^{v} \binom{v}{\alpha} \sum_{\beta}^{v-\alpha} \binom{v-\alpha}{\beta} \sum_{u=0}^{m}(-1)^{m-u} \times$$
$$\times \binom{m}{u} i^u \sum_{\mu=0}^{\left\lfloor \frac{l-m}{2} \right\rfloor} (-1)^\mu 2^{-2\mu} \binom{l}{\mu}\binom{l-\mu}{m+\mu} \sum_{\varrho=0}^{\mu} \binom{\mu}{\varrho} \qquad , \qquad (11)$$

and $w = 2(\varrho + \alpha) + u$, $s = 2(\mu - \varrho + \beta) + m - u$, $t = 2(v - \alpha - \beta - \mu) + l - m$. Since the functions in Eq. (10) form a complete orthonormal system, it is possible to approximate any 3D object by a finite number of 3D Zernike descriptors (3DZD). For our application, let $T(\mathbf{x})$ be the tomographic reconstruction, its approximation using the Zernike basis is

$$T(\mathbf{x}) \approx \sum_{n,l,m} \Omega_{nl}^m Z_{nl}^m(\mathbf{x}), \qquad (12)$$

where $\Omega_{nl}^m$ are the 3DZD. In our implementation, we enforce the tomographic reconstruction to be fitted by Eq. (12), hence the 3DZD can be calculated by solving the equivalent linear system, after vectorizing the terms in Eq. (12). Without lack of generality, we consider the tomogram $T(\mathbf{x})$ calculated within a cube having $V$ voxels per side, such that $V^3$ is total number of voxels. Let $N$ be the Zernike fitting order, such that $n \epsilon [0, N]$. It is possibile to evaluate the number of Zernike basis functions to be generated by fixing $N$ as the sum of the first $N + 1$ triangular numbers. Let $M$ be this summation, hence the Zernike functions can be seen as 4D vector with sizes $V \times V \times V \times M$, i.e. $Z_{nl}^m(\mathbf{x}) = [Z\{1\}, Z\{2\}, ..., Z\{M\}]$ with $Z\{j\}$ having the same dimension of $T(\mathbf{x}) \, \forall \, j \epsilon [1, M]$. Therefore, the Eq. (12) can be rewritten and solved as

$$\hat{\mathbf{T}} \approx \bar{\bar{\mathbf{Z}}} \hat{\mathbf{\Omega}} \Rightarrow \hat{\mathbf{\Omega}}_D = pinv\{\bar{\bar{\mathbf{Z}}}\} \hat{\mathbf{T}}, \qquad (13)$$

where $\hat{\mathbf{T}} = vec(T)$ and $\hat{\mathbf{\Omega}}_D = vec(\Omega_D)$ with dimensions $V^3 \times 1$ and $M \times 1$, respectively, and $\bar{\bar{\mathbf{Z}}} = [vec(Z\{1\}), vec(Z\{2\}), ..., vec(Z\{M\})]$, thus having sizes $V^3 \times M$. The operator $pinv$ indicate the pseudo-inverse matrix calculation. Finally, $\hat{\mathbf{\Omega}}_D$ is the vector containing the 3DZD.

*Tomograms recovery via 3DZD: performance analysis*

It is possible to evaluate the fidelity of the tomograms recovered via 3DZD with respect to the ground truth for a fixed $M$, e.g. by calculating the RMSE between them. Notice that $\lim_{M \to \infty} RMSE = 0$. This means that one can exploit the RMSE as a stop criterion in approximating a tomogram with 3DZD. In fact, in our implementation, we have increased the number of Zernike polynomials $M$ up to reach a NRMSE lower than 1% in the case of tomographic cell phantom reconstruction in Fig.1. As result, we need to fix $N = 30$ to approximate the tomograms with that fidelity. Then, we have used this setting to compute the 3DZD fitting for the single-cell tomograms in Fig.2, achieving a NRMSE < 2% in all cases. It is important to underline that, we have used tomographic data of sizes 50x50x50 for which the above setting ensures the desired fidelity score. One can expect that for data with higher resolution, the above condition NRMSE < 1% may be achieved with $N > 30$ since tinier intracellular structures need to be shaped by higher order of Zernike polynomials.

In principle, the end users of the proposed method can select any of the many reference-based similarity metrics proposed in the literature as the stop criterion in generating Zernike polynomials. Currently, one of the most used metric is the Structural Similarity Index (SSIM), that is to say a perception-based metric able to quantify the structural information related to the processed image distortion, taking into account both luminance and contrast. Other commonly used metrics are the Peak Signal-to-Noise Ratio (PSNR), employed to quantify the reconstruction quality on an image under lossy compression, and the Mean Absolute Error (MAE) that estimates the so-called absolute error between a pair of images. In Fig.3 we replicate the same analysis of Fig.1e, thus quantifying the fidelity of the tomographic cell phantom recovery by 3DZD in the case of SSIM (Fig.3a), PSNR (Fig.3b) and MAE (Fig.3c). We have included in the top also fidelity scores in the case of $N = 30$. Notice that, the SSIM and the MAE exhibit a non-monotone trend as a function of the Zernike order, being unstable for low orders. Instead the PSNR seems to be an appropriate fidelity score, as one can expect since it is a function of the MSE.

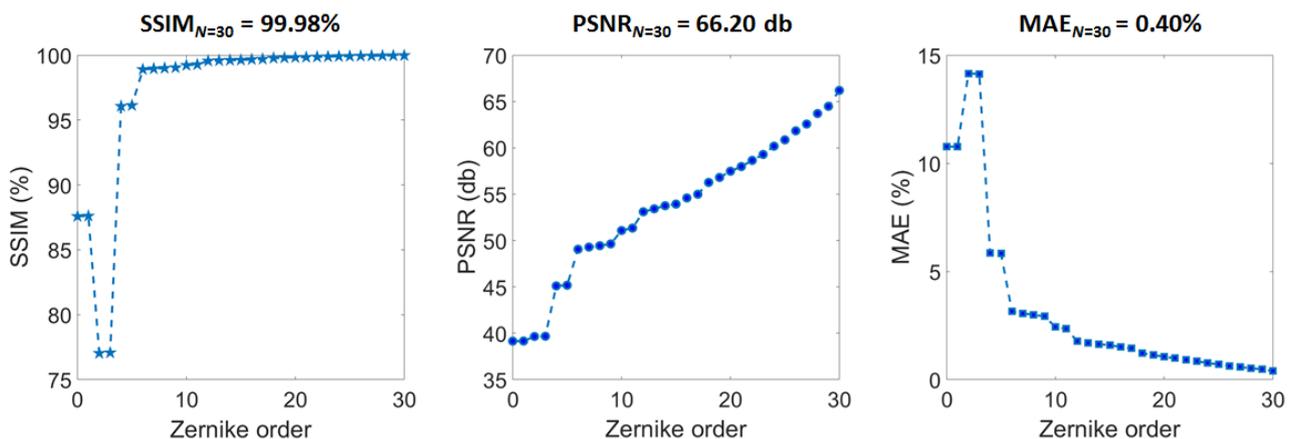

**Fig. 3. Most used reference-based similarity metrics to quantify the tomographic cell phantom recovery via 3DZD.** SSIM **(a)**, PSNR **(b)** and MAE **(c)** evaluated by increasing the Zernike order up to $N = 30$. The corresponding best scores are reported on the top.

The same performance analysis is repeated for the data in Fig. 2 a-f and summarized in the following Table 1.

|  | SSIM (%) | PSNR (db) | MAE (%) |
|---|---|---|---|
| Red blood cell – Fig.2a | 99.94 | 61.11 | 0.41 |
| White blood cell (THP-1) – Fig.2b | 99.95 | 62.61 | 0.37 |
| SKOV3 ovarian cancer cell – Fig.2c | 99.96 | 64.21 | 0.41 |
| CHP212 neuroblastoma cancer cell – Fig.2d | 99.93 | 61.59 | 0.81 |
| NIH-3T3 mouse cell – Fig.2e | 99.96 | 63.71 | 0.54 |
| Yeast cell – Fig.2f | 99.23 | 49.66 | 0.72 |

**Table 1. Fidelity scores for single-cell tomograms in Fig. 2 a-f.**

*Optical setups for tomographic data recording*

The tomograms reported in Fig. 2 have been reconstructed by using three different setups, hereinafter referred to setup A[6], setup B[9] and setup C[18]. The setup A and the setup B share the flow cytometry condition, so their layout looks similar differing only in some optical components. Instead, the setup C is based on the illumination scanning for tomographic imaging of static samples[18]. More precisely, setup A and setup B are equipped by a DH microscope in off-axis configuration based on a Mach-Zehnder interferometer (Fig. 4a). The green laser beam (Laser Quantum Torus 532) is split into a reference and an object beam by a polarizing beam splitter (PBS) cube. Two half-wave plates (HWPs) are placed in front of and behind the PBS in order to balance the ratio between the intensity of the object and reference beams while keeping the same polarization. The object beam illuminates the cells while flowing and rotating along a microfluidic channel (Microfluidic ChipShop 10000107 - 1000μm×200μm) thanks to the push of an automatic syringe pump (Syringe Pump neMESYS 290N – 7nl/s for setup A – 50nl/s for setup B), and then it is collected by a microscope objective ($MO_A$ Zeiss Plan-Apochromat 60× NA=1.2 Water immersion for setup A – $MO_B$ Zeiss Plan-Apochromat 40× NA=1.3 Oil immersion for setup B). The object beam is recombined with the reference beam by a beam splitter cube (BS) in order to generate the interference pattern digitally recorded by the camera (CCD u-eye from IDS, 2048×2048 pixels, 5.5μm pixel size for setup A - CMOS Genie Nano-CXP Cameras, 5120×5120 pixels, 4.5μm pixel size for setup B). According to the reference system in Fig. 4a, cells flow along the y-axis and rotate around the x-axis, while the holographic sequence is recorded along the z-axis (75fps for setup A - 30fps for setup B). In case of setup C (Fig. 4b), the optical system used a diode pumped solid state 532nm laser. The laser beam was first spatially filtered using a pinhole spatial filter. A beam splitter was used to separate the input beam into a sample beam and a reference beam. The sample beam was directed onto the sample at different angles of incidence using a reflective LCOS spatial light modulator (SLM) (Holoeye) with a pixel size of 8μm and resolution of 1080×1920 pixels. Different illumination angles were obtained by projecting blazed gratings on the SLM. In the experiments presented here, a blazed grating with a period of 25 pixels (200μm) was circularly rotated with a resolution of 1 projection per degree for total projections. Two 4f systems between the SLM and the sample permitted filtering of higher orders reflected from the SLM (due to limited fill factor and efficiency of the device) as well as magnification of the SLM projections onto the sample. Using a 100× oil immersion objective lens with NA 1.4 (Olympus), the incident angle on the sample corresponding to the 200μm grating was 35°. The magnification of the illumination side was defined by the 4f systems we used before the sample. A third 4f system after the sample includes a 100× oil immersion objective lens with NA 1.45 (Olympus). The sample and reference beams were collected on a second beam splitter and projected onto a scientific CMOS camera (Neo, Andor) with a pixel size of 6.5μm and resolution of 2150×2650 pixels.

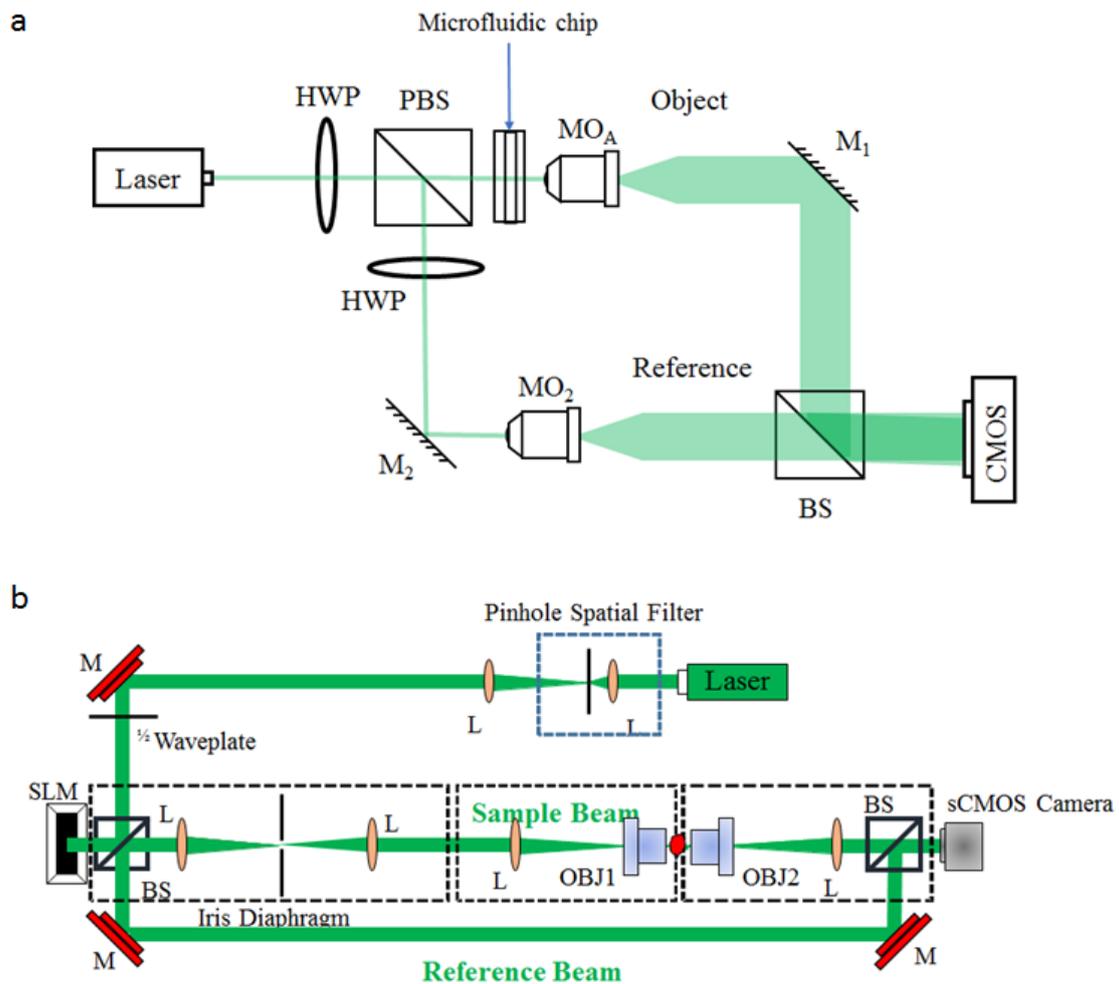

**Fig. 4. Sketches of optical setups used for tomographic imaging of cells.** Imaging systems used for recording tomographic data **(a)** in flow cytometry condition and **(b)** by illumination scanning based arrangement.

*Processing pipeline for tomograms reconstructions*

To reconstruct the 3D RI tomogram in setups A and B, a region of interest (ROI) is selected around the cell while flowing along the microfluidic channel within the recorded holographic sequence. After filtering out undesired diffraction orders terms from the Fourier spectrum of the recorded hologram, the resulting complex wavefront is propagated along the optical z-axis by means of the Angular Spectrum method in order to refocus the imaged cell[21]. In particular, the in-focus distance is computed by minimizing the Tamura Coefficient (TC) of the amplitude of the propagated complex wavefront[22]. The wrapped phase map (WPM) is extracted by computing the argument of the in-focus complex field. The phase aberrations are compensated with a fitting-based processing[23] (setup A) or by subtracting the WPM of a reference hologram recorded without the sample in the imaged FoV[24] (setup B). The resulting aberration-free WPM is then denoised and unwrapped by the PEARLS algorithm[25] in the case of setup A, while, for setup B, the WPM is denoised through the two-dimensional windowed Fourier transform filtering[26] and unwrapped through the PUMA algorithm[28]. The final quantitative phase map (QPM) is then obtained by centring the cell in its ROI after computing its transversal x-y position through the weighted centroid method[22]. In order to implement a tomographic algorithm, the corresponding viewing angle must be coupled to each QPM of the flowing and rolling cell. However, for the in-flow configurations of setups A and B, this information is not a priori known, therefore it must be estimated from the data. In particular, the unknown rotation angles of the flowing red blood cell in Fig.2a are computed by means of the Zernike coefficient of the biolensing modelling of cell[6,17] (setup A). Instead, rotation angles of cells in Figs.2 (b-e) have been calculated by using an ad-hoc phase image similarity metric[28] (setup B). Finally, the tomographic reconstruction of the cell is computed by feeding the FBP algorithm through the pair made of the QPMs and the corresponding rotation angles[6,28]. As regards the setup C, the complex field is retrieved from the intensity images by taking the Fourier transform of the holographic intensity images, selecting the order of interest, and then taking an inverse Fourier transform. This is repeated for all the projections. The Fourier diffraction theory is then used to reconstruct the 3D RI reconstruction based on the Rytov approximation. The latter serves as initial guess to the LT algorithm, in which it is updated iteratively by minimizing an error function between the experimentally measured field and the field retrieved from the forward model based on Lippmann-Schwinger equation[19]. The main steps for the tomographic reconstructions are summarized in Fig. 5.

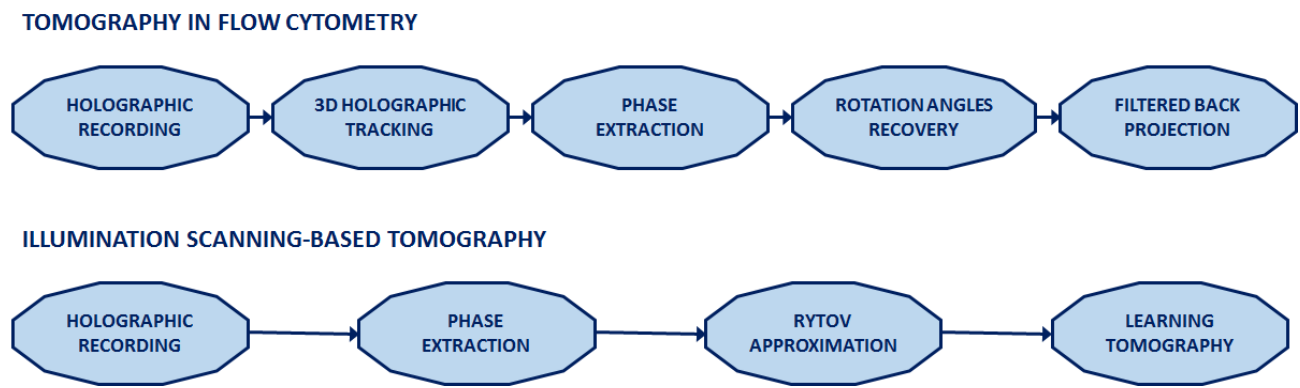

**Fig. 5. Computational steps to obtain the tomographic reconstructions.** Processing pipelines for the tomograms recovery for data collected within the recording systems in flow cytometry conditions (up row) and illumination scanning for static samples (bottom row).

# Acknowledgements

This research has been partially supported by the project MUR-PRIN 2017 Morphological Biomarkers for early diagnosis in Oncology (MORFEO) Prot. 2017N7R2C.

# Data availability

The data and the code that support the findings of this study are available from the authors upon reasonable request.